\documentclass[12pt,preprint]{emulateapj}
\def\apj{{ApJ}}

\def\be{\begin{equation}}
\def\ee{\end{equation}}
\def\bea{\begin{eqnarray}}
\def\eea{\end{eqnarray}}

\usepackage{graphicx}

\begin{document}

\title{High energy emission of GRB 130427A: evidence for inverse Compton radiation}

\author{Yi-Zhong Fan\altaffilmark{1}, P.H.T.~Tam\altaffilmark{2}, Fu-Wen Zhang\altaffilmark{1}, Yun-Feng Liang\altaffilmark{3}, Hao-Ning He\altaffilmark{1}, Bei Zhou\altaffilmark{1,4}, Rui-Zhi Yang\altaffilmark{1,4}, Zhi-Ping Jin\altaffilmark{1}, and Da-Ming Wei\altaffilmark{1}}
\affil{$^1$ Key Laboratory of Dark Matter and Space Astronomy, Purple Mountain Observatory, Chinese Academy of Science,
Nanjing, 210008, China.}
\affil{$^2$ Institute of Astronomy and Department of Physics, National Tsing Hua University, Hsinchu 30013, Taiwan.}
\affil{$^3$ Department of Physics, Guangxi University, Guangxi 530004, China.}
\affil{$^4$ Graduate University of Chinese Academy of Sciences,
Yuquan Road 19, Beijing, 100049, China.}

\begin{abstract}
A nearby super-luminous burst GRB 130427A was simultaneously detected by six $\gamma$-ray space telescopes ({\it Swift}, Fermi-GBM/LAT, Konus-Wind,  SPI-ACS/INTEGRAL, AGILE and RHESSI) and by three RAPTOR full-sky persistent monitors. The isotropic $\gamma-$ray energy release is of $\sim 10^{54}$ erg, rendering it the most powerful explosion among the GRBs with a redshift $z\leq 0.5$.
The emission above 100 MeV lasted about one day and four photons are at energies greater than 40 GeV. We show that the count rate of 100 MeV-100 GeV emission may be mainly accounted for by the forward shock synchrotron radiation and the inverse Compton radiation likely dominates at GeV-TeV energies. In particular, an inverse Compton radiation origin is favored for the $\sim (95.3,~47.3,~41.4,~38.5,~32)$ GeV photons arriving at $t\sim (243,~256.3,~610.6,~3409.8,~34366.2)$ s after the trigger of Fermi-GBM. Interestingly, the external-inverse-Compton-scattering of the prompt emission (the second episode, i.e., $t\sim 120-260$ s) by the forward-shock-accelerated electrons is expected to produce a few  $\gamma-$rays at energies above $10$ GeV, while five were detected in the same time interval. A possible unified model for the prompt soft $\gamma-$ray, optical and GeV emission of GRB 130427A, GRB 080319B and GRB 090902B is outlined. Implication of the null detection of $>1$ TeV neutrinos from GRB 130427A by IceCube is discussed.
\end{abstract}

\keywords{Gamma rays: general---Radiation mechanisms:
non-thermal}

\setlength{\parindent}{.25in}

\section{Introduction}
The high-energy ($\geq 100$ MeV) emission properties of Gamma-ray Bursts (GRBs) can help us to
better understand the physical composition of the GRB outflow,
the radiation mechanisms, and possibly also the underlying physical
processes shaping the early afterglow (see Fan \& Piran 2008 and Zhang \& M\'esz\'aros 2004 for reviews). For example,
the inverse Compton radiation from GRB forward shock can extend to the very
high energy ($\epsilon_\gamma >50~{\rm GeV}$) range \citep{Dermer2001,Sari2001,Zhang2001,Fan08}, while the synchrotron radiation can only give rise to emission up to $\sim 10~{\rm GeV}~(\Gamma/100)(1+z)^{-1}$, where $\Gamma$ is the bulk Lorentz factor of the GRB blast wave and drops with time quickly. Hence at $t>10^{2}$ s, usually we do not expect tens-GeV $\gamma-$rays coming from the synchrotron radiation of the forward shock. Therefore the detection of very high energy emission of GRBs at a fairly early time can impose a very tight constraint on the radiation mechanism. However, the very high energy photons are rare and are attenuated by the cosmic infrared/optical background before reaching us. Xue et al. (2009) investigated the detection prospect of very high energy emission of GRBs and found out that with current ground-based Cherenkov detectors, only for those very bright and nearby bursts like GRB 030329, detection of very high energy photons is possible under favorable observing conditions and for a delayed observation time of $\leq 10$ hr. Very bright and nearby bursts are very rare and for the ground-based detectors the observation conditions are not under control. That's why so far no positive detection of very high energy emission from GRBs by the ground-based Cherenkov detectors has been reported, yet (e.g., Albert et al. 2007; Horan et al. 2007; Aharonian et al. 2009; Jarvis et al. 2010; Acciari et al. 2011)\footnote{HAWC
observed GRB 130427A. However, at the time of the GBM trigger, the elevation of the burst in
HAWC's field of view was only 33.13 degrees and setting. At such an elevation the
sensitivity of HAWC is more than 2 orders of magnitude
poorer than near the zenith \citep{Lennarz2013}. Hence the non-detection is not surprising.}.

In comparison with the ground-based Cherenkov detectors, the space telescopes such as EGRET onboard CGRO, GRID onboard AGILE, and the Large Area Telescope (LAT) onboard the Fermi satellite have a much smaller effective area. However these telescopes have a low energy threshold $\sim$ tens MeV and can monitor the high energy emission since the trigger of some GRBs when the high energy emission flux is expected to be much higher than that at late times. Since 1994, tens GRBs with high energy emission have been reported. In the pre-Fermi-LAT era, the record of the most energetic $\gamma-$ray from GRBs is the $\sim 18$ GeV photon following GRB 940217 (Hurley et al. 1994). The most energetic photon detected by Fermi-LAT till March 2013 is the $33.4$ GeV $\gamma-$ray from GRB 090902B at a redshift $z=1.822$ \citep{Palma09,Abdo2013}. Such a record has been broken by GRB 130427A, a burst simultaneously detected by {\it Swift} \citep{Maselli2013}, Fermi Gamma-Ray Telescope \citep{Zhu2013,Kienlin2013}, Konus-Wind \citep{Golenetskii2013}, SPI-ACS/INTEGRAL \citep{Pozanenko2013}, AGILE \citep{Verrecchia2013}, RHESSI \citep{Smith2013} and three RAPTOR full-sky persistent monitors \citep{Wren2013}. The highest energy LAT photon has an energy of $> 90$ GeV \citep{Zhu2013}. The redshift of this burst was measured to be $0.3399\pm 0.0002$ \citep{Flores2013,Levan2013,Xu2013} and a bright supernova SN 2013cq was identified \citep{Xu2013}. Its isotropic energy release in the energy range $20-10^{4}$ keV is $E_{\rm \gamma,iso} \sim 8.5\times 10^{53}$ erg \citep{Golenetskii2013}, rendering it the most energetic one among the GRBs with a redshift $z\leq 0.5$ detected so far. As shown in Tab.\ref{Tab:low-z}, GRB 130427A is far more energetic than all other low redshift GRBs.

\begin{table*}[t!]
\caption {The observational properties of low redshift GRBs ($z<0.5$)}
\begin{tabular}{lllll}
\hline
\hline

GRB & $z$  & $E_{\rm peak}$ \tablenotemark{a} &$E_{\rm \gamma,iso}$       &  Ref.\tablenotemark{b}\\

   &         &   (keV)       & ($10^{51}$ erg)           &     \\
  \hline

990712	&0.434	&93$\pm$15	   &6.7$\pm$1.3		  & 1   \\		
980425  &0.0085 &122$\pm$17    &$9\times10^{-4}$  & 2   \\					
010921	& 0.45	&129$\pm$26	   & 9.5$\pm$1        & 1   \\
011121	& 0.36	&1060$\pm$265  & 78$\pm$21	      & 1   \\
020819B	& 0.41	&70$\pm$21	   & 6.8$\pm$1.7      & 1   \\
020903	& 0.25	&3.37$\pm$1.79 &0.024$\pm$0.006   & 1   \\
030329	& 0.168	& 100$\pm$23   &15$\pm$3          & 1   \\
031203  &0.105  &$>190$        &0.17               & 2   \\
040701  &0.215  &$<6$          &0.08$\pm$0.02      & 3   \\
050509B &0.226 & $\sim$101          & $\sim 9\times10^{-3}$  &4      \\
050709	&0.1606	&97.4$\pm$11.6  &0.033$\pm$0.001 & 1   \\
050724	&0.258    &$\sim$126        & $\sim$0.35       & 4   \\
050826	&0.297    &$\sim$441        & $\sim$0.33       & 4   \\
051117B	&0.481    & $\sim$107        &$\sim$13                & 4   \\
060218	& 0.0331 &4.9$\pm$0.3	&0.053$\pm$0.003  & 3   \\
060505	&0.089	 &$>160$        & 0.03$\pm$0.01   & 3   \\
060614	&0.125	 &55$\pm$45	    &2.5$\pm$1        & 1   \\
061006	&0.4377	 &955$\pm$267	&2$\pm$	0.3       & 1   \\
061021	&0.346	 &1046$\pm$485	&4.6$\pm$0.8      &1    \\
061210	&0.4095  &$\sim$767       &$\sim$0.91     & 4   \\
071227	&0.383	 &1384$\pm$277	&1$\pm$	0.2      &1    \\
080905A	&0.1218  &$\sim$503         & $\sim$0.55      & 4   \\
090417B	&0.345	 & --	        &$>$6.3           &5    \\
091127	&0.49	 &51$\pm$1.5	& 16.1$\pm$0.3    &1    \\
100206A &0.4068  &618$\pm$103  &0.62$\pm$0.03      & 6 \\
100316D &0.0591   &18$^{+3}_{-2}$         &0.06   &2   \\
111211A	&0.478   & -             & $\sim$11              & 4   \\
120422A &0.283   & $\sim$ 53        &0.045         & 2 \\
120714B	&0.3984  &$\sim$99          & 7.95$\pm$0.09   & 4  \\
130427A	&0.3399	 &1378$\pm$11	& $\sim 850$      &7\\
\hline
\hline
\end{tabular}
\label{Tab:low-z}
\begin{list}{}{}
\item[$^{(a)}$] The peak energy of the prompt emission in the burst frame.
\item[$^{(b)}$] References: (1) Zhang et al. 2012b and references therein, (2) Zhang et al. 2012a, (3)Amati et al. 2007,  (4)  Butler et al. 2007 (http://butler.lab.asu.edu/swift/), (5) Holland et al. 2010, (6) von Kienlin 2010, (7) Golenetskii et al. 2013.
\end{list}
\end{table*}

The prompt emission of GRB 130427A lasted 
a few hundred seconds and overlapped with the forward shock region significantly. The forward shock protons and electrons are also cooled by the prompt emission and high energy neutrinos and $\gamma-$rays are powered by the ultra-high energy protons interacting with the prompt $\gamma-$rays and by the electrons inverse-Compton-scattering off the prompt emission \citep{Fan05a,Fan05b}. As a result of the Klein-Nishina suppression, the forward shock electrons are mainly cooled by the X-ray photons (Fan et al. 2005b, Wang et al. 2006). Therefore in section 2 we analyze the Swift BAT data and then extrapolate the 0.3-10 keV flux of the prompt emission. The 100MeV-100 GeV photon flux and the arrival time of the $>1$ GeV photons of GRB 130427A are also presented. In section 3 we examine the models of the long-lasting GeV emission. In section 4 we discuss the physical origin of the prompt emission. In section 5 we summarize our work with some discussion.

\section{The prompt 0.3-10 keV emission and the 100MeV-100GeV emission}
In this work $T_0$ denotes the trigger of Fermi-GBM at 07:47:06.42 UT on 27 April 2013 \citep{Kienlin2013}. The {\it Swift} BAT was executing a pre-planned slew, so it was not triggered on time. But in the BAT lightcurve the main large peak started about $50$ s before its trigger \citep{Maselli2013,Barthelmy2013}, consistent with the observations of Fermi-GBM, Konus-Wind, SPI-ACS/INTEGRAL, AGILE and RHESSI.

\begin{figure}
\begin{picture}(0,210)
\put(0,0){\includegraphics{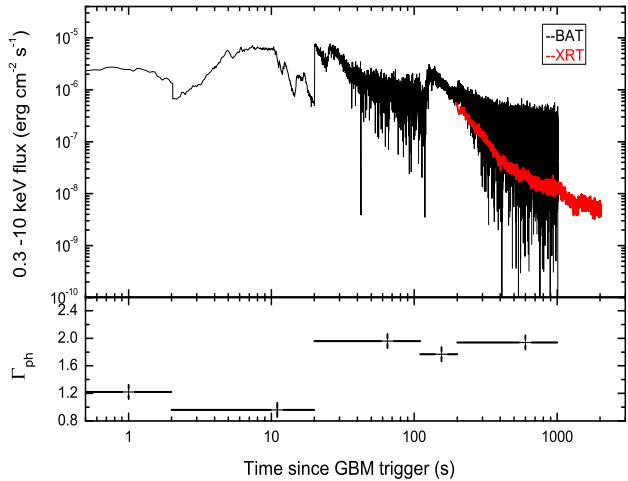}}
\end{picture}
\caption{Upper panel: prompt $0.3-10$ keV emission of GRB 130427A extrapolated from the BAT data. The red line is the XRT lightcurve taken from http://www.swift.ac.uk/xrt\_curves/00554620/ (Evans et al. 2009). Lower panel: the photon index ($\Gamma_{\rm ph}$) of the prompt emission detected by BAT in different intervals.}
\label{fig:130427A-XRT}
\end{figure}

\emph{{The prompt 0.3-10 keV emission.}}
 The BAT quicklook data were analyzed using the standard BAT analysis software distributed within HEASOFT 6.13 and the latest calibration files. The BAT ground-calculated position is RA= 173.150, Dec= 27.706 deg. Mask-tagged BAT count-rate light curve was extracted in the standard 15-150 keV energy bands, and converted to 15-150 keV flux with the energy conversion factor inferred from the spectral fitting in different time-intervals shown in  Fig.\ref{fig:130427A-XRT}, where a simple power-law model is adopted. Assuming the spectrum is unchanged in the energy range $0.3-10$ keV, the prompt X-ray emission lightcurve is extrapolated (see Fig.\ref{fig:130427A-XRT}).

\emph{{The 100 MeV - 100 GeV emission.}}
The first reports of the LAT emission were made by Zhu et al. (2013a, b). To better understand the high-energy emission we analyzed the LAT data that are available at the Fermi Science Support Center\footnote{\url{http://fermi.gsfc.nasa.gov/ssc/}}, using the Fermi Science Tools v9r27p1 package. Events of energies between 100~MeV and 100~GeV were used. To reduce the contamination from Earth albedo $\gamma$-rays, we excluded events with zenith angles greater than 100$^\circ$. Since we focused our work in the extended LAT emission that lasts for nearly a day, events classified as ``P7SOURCE'' and the instrument response functions ``P7SOURCE\_V6'' were used.

Events from a region-of-interest (ROI) of a 20$^\circ$-radius circular region centered on the enhanced XRT position of GRB~130427A (GCN 14467) were analyzed using unbinned likelihood analyzes. The Galactic diffuse emission (gal\_2yearp7v6\_v0.fits) and the isotropic diffuse component (iso\_p7v6source.txt), as well as sources in the second Fermi catalog were included in the background model. However, it was shown that an isotropic component is enough to describe the background photons in the time bins before $T_0+1000$~s, due to the dominance of the GRB emission over other sources in the ROI during these short-duration intervals.

\begin{figure}
\begin{picture}(0,220)
\put(0,0){\includegraphics{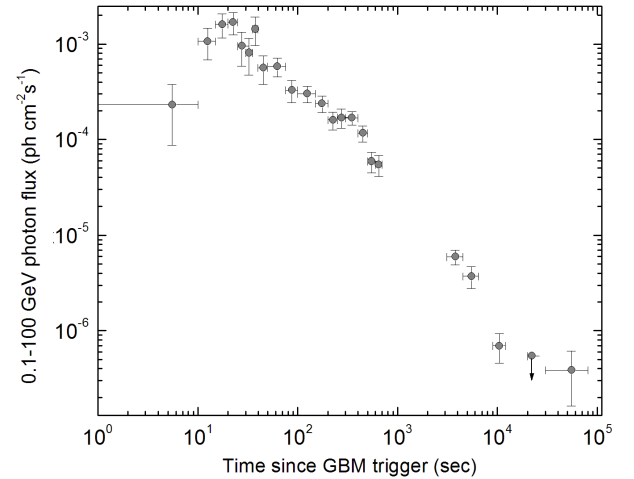}}
\end{picture}
\caption{The 0.1-100 GeV photon flux of GRB 130427A. At $t>100$ s, the count rate can be approximated by $\dot{{\cal N}}=3\times 10^{-4}~{\rm cm^{-2}~s^{-1}}~(t/100~{\rm s})^{-1.2}$.}
\label{fig:130427A-GeV}
\end{figure}

We proceeded to construct a light curve in the energy interval 100~MeV to 100~GeV, using unbinned likelihood analyzes of each time bin. The light curve is shown in Fig.\ref{fig:130427A-GeV}. Due to the brightness of the GRB, intervals were as short as 5--10 seconds in the early times, e.g., before $T_0+50$~s. As already noted in \citet{Zhu2013b}, there is a possible break at around $T_0+500$~s. Spectral analyzes on two time intervals: $T_0$ to $T_0+138$~s and $T_0+138$~s to $T_0+70$~ks were carried out. Spectral analyzes for the 0.1-100 GeV data during two time intervals: (I) T0 to T0 +138 s and (II) T0 + 138 s to T0 + 70 ks is were carried out. Assuming single power laws for the spectra of the GRB, we found that $\Gamma_\mathrm{0.1-100 GeV}=-1.9\pm0.1$ during period (I) and $\Gamma_\mathrm{0.1-100 GeV}=-2.1\pm0.1$ during period (II). Therefore, within uncertainties, the spectrum remains unchanged between the two periods. A more in-depth LAT analysis of GRB 130427A has been done by Tam et al. (2013).

\begin{figure}
\begin{picture}(0,240)
\put(0,0){\includegraphics{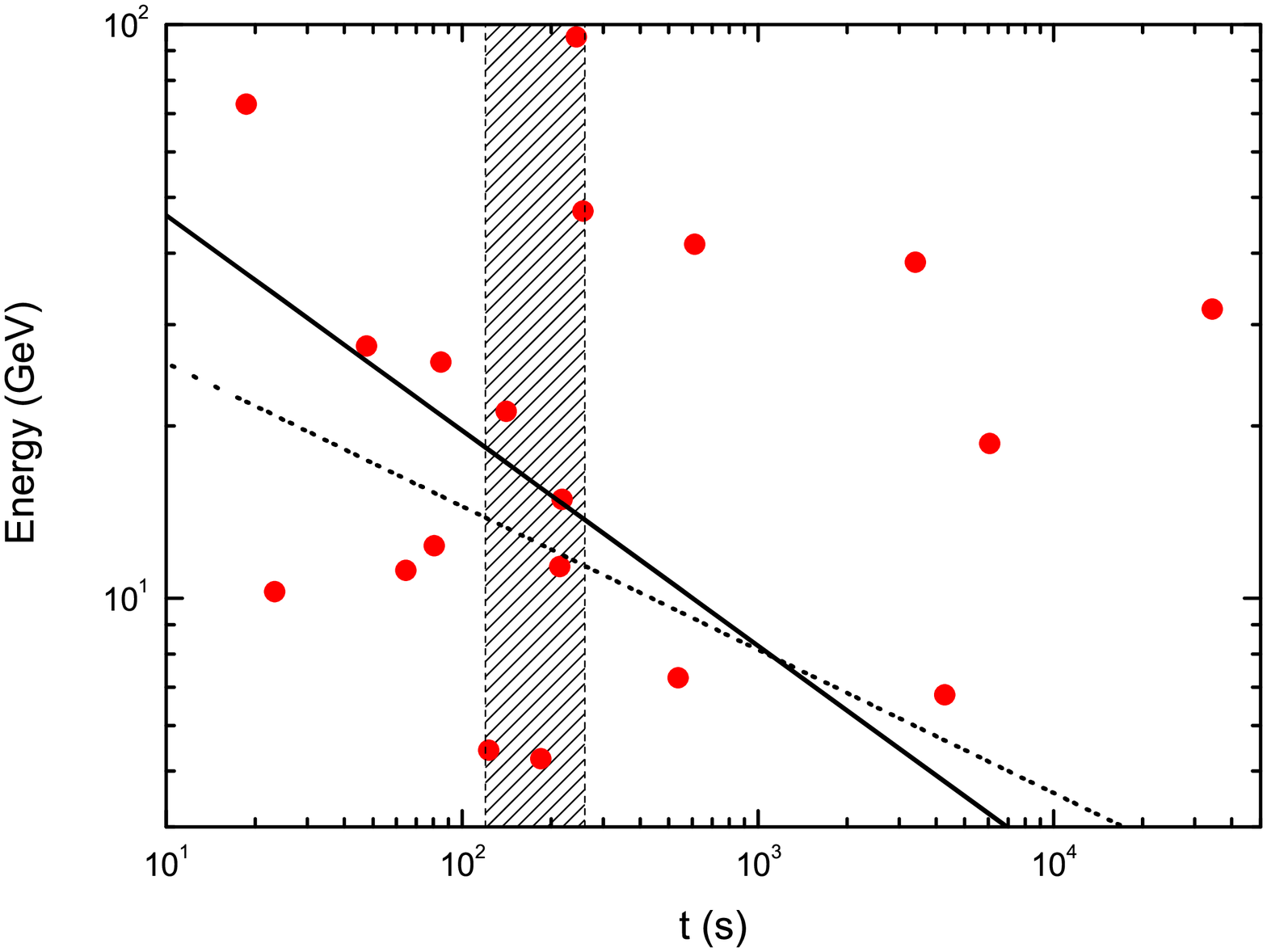}}
\end{picture}
\caption{The arrival time of the $\geq 5$ GeV photons as well as the expected maximal synchrotron radiation frequency of the forward shock emission given by eqs.(\ref{eq:Syn_limit}). The solid line is for the ISM case ($n=0.01~{\rm cm^{-3}}$) while the dotted line is for the wind medium ($A_*=0.01$). The isotropic-equivalent kinetic energy of the GRB ejecta is taken to be $10^{54}$ erg. The shaded area represents the time interval of the second episode of the prompt emission, in which strong GeV-TeV radiation of the forward shock electrons caused by inverse-Compton-scattering off prompt photons (Fan et al. 2005a, 2005b) is expected.}
\label{fig:130427A-GeVArrival}
\end{figure}

In Fig.\ref{fig:130427A-GeVArrival} we also present the photon energies and arrival times of the photons (to increase the photon statistics, a looser selection criterium: ``P7TRANSIENT'' is employed here) above 5 GeV since the trigger of Fermi GBM. The 95\% contamination angle of LAT at 5 GeV is about $1^\circ$, that's why in our plot only the photons within the $1^\circ$ aperture have been taken into account.

\bigskip

\bigskip

\bigskip

\section{Physical origins of the GeV emission}
The circum-burst medium can be either interstellar medium (ISM) or stellar wind, which may be hard to reliably distinguish. In view of such an uncertainty we discuss both scenarios. In the ISM model, the number density of the medium ($n$) is taken to be a constant while for the wind medium we take $n_{\rm w}=3\times 10^{35}A_* R^{-2}$, where $A_*$ is the dimensionless wind parameter and $R$ is the radius of the forward shock. The fact that the forward shock X-ray emission may be in slow cooling phase at a time $t\sim 0.1$ day favors a low density circum-burst medium \citep{Laskar2013}. That's why in the following investigation we normalize $n$ to $0.01~{\rm cm^{-3}}$ and $A_*$ to $0.01$.

The absence of a clear jet break up to $t>80$ days suggests a half-opening angle of \citep{Piran99,Mesz02,Zhang04}
\begin{equation}
\theta_{\rm j}>
\left\{%
\begin{array}{ll}
    0.23~E_{\rm k,54}^{-1/8}n_{-2}^{1/8}(t/20~{\rm d})^{3/8}({1+z\over 1.34})^{-3/8}, & \hbox{ISM;} \\
    0.08~E_{\rm k,54}^{-1/4}A_{*,-2}^{1/4}(t/80~{\rm d})^{1/4}({1+z \over 1.34})^{-1/4}, & \hbox{wind;} \\
\end{array}%
\right.
\end{equation}
and the intrinsic $\gamma-$ray energy release of the ejecta is
\begin{equation}
E_{\rm \gamma,jet,51}>
\left\{%
\begin{array}{ll}
    25~E_{\rm \gamma,iso,53.93}E_{\rm k,54}^{-1/4}n_{-2}^{1/4}
({t \over 20~{\rm d}})^{3/4}, & \hbox{ISM;} \\
     3~E_{\rm \gamma,iso,53.93}E_{\rm k,54}^{-1/2}A_{*,-2}^{1/2}({t\over 20~{\rm d}})^{1/2}, & \hbox{wind;} \\
\end{array}%
\right.
\end{equation}
where $E_{\rm k}$ is the isotropic-equivalent kinetic energy of the GRB ejecta. Here and throughout this text, the convention $Q_{\rm x} =Q/10^{\rm x}$ has been adopted in CGS units except for specific notation.

\citet{Laskar2013} and \citet{Perley2013} adopted the wind medium model to interpret the multi-wavelength afterglow data and claimed the discovery of reverse shock optical/radio radiation components. In their modeling the bulk Lorentz factor of the reverse-shock-heated ejecta is required to drop with time as $\Gamma \sim 130(t/200~{\rm s})^{-5/11}$ (in the reverse shock theory, the drop can not be quicker than $(t/200~{\rm s})^{-1/3}$; S. Kobayshi and Y. C. Zou 2013, private communication). If correct, the absence of the jet effect in the radio afterglow data in at least 10 days suggests a very wide half-opening angle $\theta_{\rm j}>0.35$ and then an intrinsic $\gamma-$ray energy release $E_{\rm \gamma,jet}>5\times 10^{52}$ erg. Such a huge $E_{\rm \gamma,jet,51}$ is above the maximal kinetic energy of a quickly-rotating neutron star and points towards a black hole central engine.

\citet{Liu2013} adopted a jet break at $t\sim 0.6$ day to interpret the afterglow model. In such a scenario $E_{\rm \gamma,jet} \sim~ 10^{51}$ erg and the early (i.e., $t<0.6$ day) X-ray afterglow emission should be attributed to the prolonged activity of the central engine which likely plays an important role in producing GeV-TeV emission via EIC process.

\subsection{The role of synchrotron radiation}\label{sec:2.1}
The forward shock synchrotron radiation of some extremely bright GRBs can play the dominant role in producing $<10$ GeV afterglow emission, as found by Zou et al. (2009, Fig.3 therein) in modeling GRB 080319B. After the first official release of the Fermi-LAT data on GRBs, the synchrotron radiation has been found to be necessary for interpreting the high energy afterglow data (e.g., Kumar \& Barniol Duran 2009; Gao et al. 2009; Ghisellini et al. 2010; He et al. 2011; Ackermann et al. 2013; c.f., Tam et al. 2012).

To produce the $\geq 0.1$ GeV afterglow emission, the synchrotron-radiating electrons should have a random Lorentz factor larger than
\begin{eqnarray}
\bar{\gamma}_{\rm e} &\sim  & 10^{7}\Gamma_2^{-1/2}B^{-1/2}(1+z)^{1/2}.
\end{eqnarray}
In the case of ISM, the bulk Lorentz factor of the forward shock reads $\Gamma\sim 270~E_{\rm k,54}^{1/8}n_{-2}^{-1/8}t_2^{-3/8}[(1+z)/1.34]^{3/8}$ and $B\sim 0.5~{\rm Gauss}~\epsilon_{\rm B,-2}^{1/2}E_{\rm k,54}^{1/8}n_{-2}^{3/8}t_2^{-3/8}[(1+z)/1.34]^{3/8}$ is the strength of shock-generated magnetic field (Piran 1999), and $\epsilon_{\rm B}$ ($\epsilon_{\rm e}$) is the fraction of the shock energy given to the magnetic field (electrons). In the wind medium, we have $\Gamma \sim 200~E_{\rm k,54}^{1/4}A_{*,-2}^{-1/4}t_2^{-1/4}[(1+z)/1.34]^{1/4}$ and $B~\sim 0.8~{\rm Gauss}~\epsilon_{\rm B,-2}^{1/2}A_{*,-2}^{3/4}E_{\rm k,54}^{1/4}t_2^{-1/4}[(1+z)/1.34]^{1/4}$ (Dai \& Lu 1998).

The inverse Compton cooling is in the Klein-Nishina regime and thus get suppressed if the seed photons are more energetic than
\begin{eqnarray}
\bar{\epsilon}_{\rm s} &\sim & m_{\rm e}c^2 \Gamma/4 \bar{\gamma}_{\rm e} \nonumber\\
&\sim &\left\{%
\begin{array}{ll}
    3.5~{\rm eV}~E_{\rm k,54}^{1/4}\epsilon_{\rm B,-2}^{1/4}t_2^{-3/4}({1+z\over 1.34})^{1/4}, & \hbox{ISM;} \\
     3~{\rm eV}~E_{\rm k,54}^{1/2}\epsilon_{\rm B,-2}^{1/4}t_2^{-1/2}, & \hbox{wind;} \\
\end{array}%
\right.
\end{eqnarray}
In most cases except in the presence of a giant optical flare, the power released in the energy range $\leq \bar{\epsilon}_{\rm s}$ of afterglow emission or late prompt emission powered by the extended activity of the central engine is expected to be (well) below that of magnetic field in the forward shock region $\sim 10^{50}~{\rm erg~s^{-1}}~\epsilon_{\rm B,-2}E_{\rm k,54}(1+z)/t_2$. We hence conclude that usually the electrons generating GeV synchrotron emission do not suffer from sizable inverse Compton cooling.

The electrons producing GeV synchrotron radiation are in fast cooling. The faction ($f$) of the total electron energy given to such extremely energetic electrons can be estimated as $f\approx (\bar{\gamma}_{\rm e}^{2-p}-\gamma_{\rm M}^{2-p})/({\gamma}_{\rm m}^{2-p}-\gamma_{\rm M}^{2-p})$, where the shock-accelerated electrons are assumed to have an initial distribution $dn/d\gamma_{\rm e}\propto \gamma_{\rm e}^{-p}$ for $\gamma_{\rm m}<\gamma_{\rm e}<\gamma_{\rm M}$, the maximal random Lorentz factor of the shock-accelerated electrons is limited by their energy loss via synchrotron radiation and is estimated by $\gamma_{\rm M} \sim 10^{8}~B^{-1/2}$ (Cheng \& Wei 1996), and
\begin{eqnarray}
\gamma_{\rm m} \sim \left\{%
\begin{array}{ll}
    8000~E_{\rm k,54}^{1/8}n_{-2}^{-1/8}t_2^{-3/8}C_{\rm p}\epsilon_{\rm e,-1}({1+z\over 1.34})^{3/8}, & \hbox{ISM;} \\
     6500~E_{\rm k,54}^{1/4}A_{*,-2}^{-1/4}t_2^{-1/4}C_{\rm p}\epsilon_{\rm e,-1}({1+z\over 1.34})^{1/4}, & \hbox{wind;} \\
\end{array}%
\right.
\end{eqnarray}
where $C_{\rm p}\equiv {6(p-2)\over (p-1)}$.

The luminosity of GeV synchrotron emission can be estimated by
\begin{eqnarray}
L_{\rm GeV,syn} &\sim & f \epsilon_{\rm e}E_{\rm k}(1+z)/t \nonumber\\
&\sim & 1.3\times 10^{50}~{\rm erg~s^{-1}}~f_{-1}\epsilon_{\rm e,-1}E_{\rm k,54}t_2^{-1}({1+z \over 1.34}).
\end{eqnarray}

Let's estimate the count rate. The averaged energy of the synchrotron photons above 100 MeV is
\[<E> \sim 100~{\rm MeV}~(\Gamma_{\rm ph}-1)/(\Gamma_{\rm ph}-2)[1-(\epsilon_{\rm syn,M}/0.1~{\rm GeV})^{2-\Gamma_{\rm ph}}],\] where $\epsilon_{\rm syn,M}$ is given by eq.(\ref{eq:Syn_limit}) and $\Gamma_{\rm ph}=(p+2)/2$ is the photon spectral index. For GRB 130427A, the X-ray and optical afterglow emission suggest that $p\sim 2.2$ and $\Gamma_{\rm ph} \sim 2.1$. We then have $<E> \sim 0.2-0.4$ GeV for $\epsilon_{\rm syn,M} \sim 1-10$ GeV. Hence the count rate of the GeV synchrotron radiation is expected to be
\begin{eqnarray}
\dot{N} &\sim & L_{\rm GeV,syn}/4\pi D_{\rm L}^2 <E> \nonumber\\
&\sim & 5\times 10^{-4}~{\rm photon~cm^{-2}~s^{-1}}~f_{-1}\epsilon_{\rm e,-1}E_{\rm k,54}t_2^{-1}({1+z\over 1.34})
\nonumber\\
&&
D_{\rm L,27.7}^{-2}({<E> \over 0.4~{\rm GeV}})^{-1}.
\label{eq:dot-N}
\end{eqnarray}
Such a rate seems to be able to account for a good fraction of the observed photon flux presented in Fig.\ref{fig:130427A-GeV} that can be  approximated by $\dot{{\cal N}}=3\times 10^{-4}~{\rm cm^{-2}~s^{-1}}~(t/100~{\rm s})^{-1.2}$.

\subsection{Limitation of the electron and proton synchrotron radiation models, the electromagnetic cascade model and the secondary positron synchrotron radiation model}
\subsubsection{The electron synchrotron radiation model}
\begin{figure}
\begin{picture}(0,260)
\put(0,0){\includegraphics{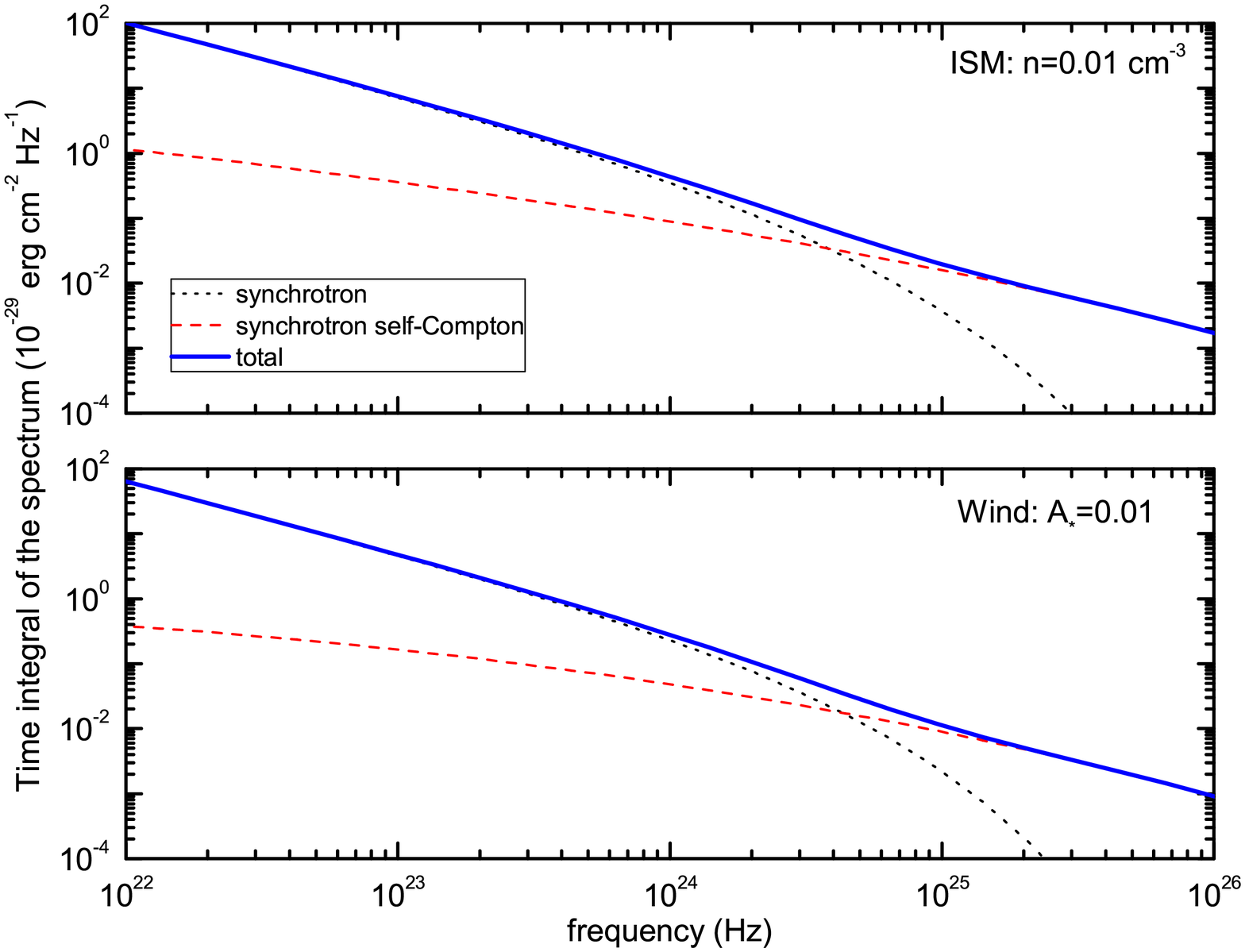}}
\end{picture}
\caption{The integral of high energy synchrotron radiation spectrum (the dotted line) and the synchrotron self-Compton radiation spectrum (the dashed line) of a {\it GRB 130427A-like} burst in the time interval $10^{2}-4\times 10^{4}$ s. The initial bulk Lorentz factor of the outflow is taken to be $300$. Other physical parameters involved in the calculation are $E_{\rm k}=10^{54}$ erg, $\epsilon_{\rm e}=0.05$, $\epsilon_{\rm B}=0.01$, $p=2.2$ and $z=0.34$. Evidently the $<10$ GeV emission is sizeably contributed by the forward shock synchrotron radiation while the inverse Compton radiation contributes at higher energies. The plots are generated from the code developed by Fan et al. (2008). Considering the rather high cooling frequency of the forward shock emission (i.e., $\nu_{\rm c}>2.4\times 10^{18}$ Hz), as suggested by the XRT spectrum, the forward shock SSC emission peaks at $\varepsilon_{\rm SSC,peak}\sim \Gamma \gamma_{\rm m}m_{\rm e}c^2$ (see eq.(23) of Zou et al. (2009) for the highly-relevant discussion). With our eq.(5),  we have $\varepsilon_{\rm SSC,peak}\sim 1~{\rm TeV}~t_2^{-3/4}$ (ISM) or $\varepsilon_{\rm SSC,peak}\sim 0.6~{\rm TeV}~t_2^{-1/2}$ (wind). On the other hand, for the burst at a redshift $z\sim 0.34$, only the $>300$ GeV photons will be significantly absorbed by the cosmic background photons and then produce cascade emission.
The total energy of these very high energy photons won't be much larger than that emitted at energies $<300$ GeV for $\varepsilon_{\rm SSC,peak}\leq 1$ TeV. Moreover, whether a sizeable fraction of the cascade emission can arrive the Earth or not strongly depends on poorly-known $B_{\rm IG}$. For $\varepsilon_\gamma \sim 1$ TeV, the cascade emission lasted $\sim 2.4\times 10^{10}~B_{\rm IG,-16}^{2}$ s (see eq.(\ref{eq:t_B})). As long as   $B_{\rm IG}>10^{-18}$ Gauss, the cascade emission caused by the $>300$ GeV SSC photons seems to be non-detectable. That is why in our plot the possible cascade emission has not been included.}
\label{fig:130427A-Prediction}
\end{figure}

At $t\sim (243,~256.3,~610.6,~3409.8,~34366.2)$ s after the trigger of Fermi-GBM, the photon with an energy $\sim (95.3,~47.3,~41.4,~38.5,~32)$ GeV was detected, respectively (see Fig.\ref{fig:130427A-GeVArrival}). The detection of such energetic $\gamma-$rays alone imposes a tight constraint on the radiation mechanism. Due to the energy loss via the synchrotron radiation, in the rest frame of the shocked medium there is an upper limit on the energy of the accelerated electrons, so is their synchrotron radiation frequency. The maximal synchrotron radiation frequency reads \citep[e.g.,][]{CW96}
\begin{eqnarray}
\epsilon_{\rm syn,M} &\sim & 100 ~{\rm MeV}~\Gamma(1+z)^{-1} \nonumber\\
&\sim & \left\{%
\begin{array}{ll}
    20~{\rm GeV}~E_{\rm k,54}^{1/8}n_{-2}^{-1/8}t_{2}^{-3/8}({1+z\over 1.34})^{-5/8}, & \hbox{ISM;} \\
     15~{\rm GeV}~E_{\rm k,54}^{1/4}A_{*,-2}^{-1/4}t_2^{-1/4}({1+z \over 1.34})^{1/4}, & \hbox{wind;} \\
\end{array}%
\label{eq:Syn_limit}
\right.
\end{eqnarray}
which is well below the energy of some photons detected in GRB 130427A (see Fig.\ref{fig:130427A-GeVArrival} for the details), suggesting that these high energy $\gamma-$rays might have an inverse Compton radiation origin. Therefore though in section \ref{sec:2.1} we have shown that the count rate of the $>100$ MeV emission may be accounted for by the synchrotron radiation of the forward shock electrons alone,  part of the high energy afterglow emission is not (see also Fig.\ref{fig:130427A-Prediction} for numerical examples).

In view of the weak dependence of $\epsilon_{\rm syn,M}$ on both $E_{\rm k}$ and $n$ or $A_*$, for almost all GRBs, the electron-synchrotron radiation origin of $>10$ GeV afterglow photons detected at $t>$ a few hundred seconds is disfavored.

\subsubsection{The proton synchrotron radiation model}
Eq.(\ref{eq:Syn_limit}) does not hold any longer if the high energy emission is from shock-accelerated protons/ions rather than electrons \citep{Razzaque2010,FP08}. However, the external forward proton radiation model has its own problem. As shown in \citet{Zhang2001}, the typical synchrotron radiation frequency of the forward shock protons read $\nu_{\rm p,m}=(\epsilon_{\rm p}/\epsilon_{\rm e})^2(m_{\rm e}/m_{\rm p})^{3}\nu_{\rm m}$, where $\nu_{\rm m}$ is the typical synchrotron radiation frequency of the forward shock electrons and $\epsilon_{\rm p}=1-\epsilon_{\rm e}-\epsilon_{\rm B}$ is the fraction of shock energy given to the protons. The maximal specific fluxes of the forward shock proton and electron synchrotron radiation are related with each other as $F_{\nu_{\rm p,max}}=(m_{\rm e}/m_{\rm p})F_{\nu_{\rm max}}$. The cooling frequency of the forward shock proton and electron synchrotron radiation are related with each other as $\nu_{\rm p,c}>(m_{\rm p}/m_{\rm e})^{6}\nu_{\rm c}$. For GRB 130427A, the hard XRT spectrum suggests that $\nu_{\rm c}\geq 10^{19}$ Hz. Hence the protons generating the GeV radiation are in the slow cooling phase with an energy distribution $dN/d\varepsilon_{\rm p} \propto \varepsilon_{\rm p}^{-p}$ and the GeV emission flux can be estimated by $F_{\nu_{\rm p,GeV}}\approx F_{\nu_{\rm p,max}}(\nu_{\rm GeV}/\nu_{\rm p,m})^{-(p-1)/2}$. As shown in \citet{Tam2013}, at energies above a few GeV, the spectrum gets hardened significantly, implying that the synchrotron radiation of the external forward shock electrons alone is unable to account for the data even if we ignore the constraint set by Eq.(\ref{eq:Syn_limit}). Such a fact suggests that $F_{\nu_{\rm p,GeV}}>F_{\nu_{\rm max}}\nu_{\rm GeV}^{-p/2}\nu_{\rm m}^{(p-1)/2}\nu_{\rm c}^{1/2}$ as long as the hard GeV emission is attributed to synchrotron radiation of the external forward shock protons, requiring that $\epsilon_{\rm p}/\epsilon_{\rm e}>(m_{\rm p}/m_{\rm e})^{(3p-1)/2(p-1)}(\nu_{\rm GeV}/\nu_{\rm c})^{-1/2(p-1)}$, i.e., $\epsilon_{\rm e}$ is much smaller than $\epsilon_{\rm p}$. As found in eq.(\ref{eq:dot-N}), $E_{\rm k} \sim 10^{54}~{\rm erg}~(\epsilon_{\rm e}/0.1)^{-1}$ is needed in the electron synchrotron radiation model. Hence in the proton synchrotron radiation model the kinetic energy of the ejecta should be $\sim 10^{53}~{\rm erg}~(m_{\rm p}/m_{\rm e})^{(3p-1)/2(p-1)}(\nu_{\rm GeV}/\nu_{\rm c})^{-1/2(p-1)}\sim 10^{58}$ erg, which is too high to be realistic.

\subsubsection{Electromagnetic cascade of TeV $\gamma-$rays}
As the $\gamma$-rays with an energy $\varepsilon_\gamma\sim 1$ TeV travel toward the
observer, a significant fraction of them will be absorbed
due to the interactions with the diffuse infrared background \citep{Plaga1995,Cheng1996,Dai2002},
yielding $e^\pm$ pairs with Lorentz factor $\gamma_{\rm e^\pm} \approx 10^{6}(\varepsilon_{\gamma}/1~{\rm TeV})$. Such ultra-relativistic electron/positron pairs will subsequently Compton scatter on the ambient cosmic microwave background
(CMB) photons, and boost them to the energy $h\nu_{\rm c,ic}\approx 0.63(1 + z)(\varepsilon_\gamma/1~{\rm TeV})^2$ GeV. Usually the duration of the resulting GeV emission is expected to be determined by the deflection
of the pairs by the intergalactic magnetic fields $B_{\rm IG}$, i.e.  \citep{Dai2002,FP08},
\begin{equation}
\Delta t_{\rm B} \sim 6.1\times 10^{11}~{\rm s}~({\varepsilon_\gamma \over 1~{\rm TeV}})^{-5}({B_{\rm IG}\over 10^{-16}~{\rm Gauss}})^{2}(1+z)^{-11}.
\label{eq:t_B}
\end{equation}
To interpret the $\sim 32$ GeV photon detected at $t\sim 3.4\times 10^{4}$ s, we need $\varepsilon_\gamma \sim 6$ TeV and $B_{\rm IG} \lesssim 3.3\times 10^{-17}$ Gauss. The origin of the $\geq 6$ TeV seed photons is unclear. Moreover, in the electromagnetic cascade scenario, as long as the magnetic deflection time governs the duration of the GeV emission, it is straightforward to show that the GeV emission should drop with the time as $t^{-1/2}$, inconsistent with the data. Such a result holds as long as the TeV seed photons have an initial duration $\ll \Delta t_{\rm B}$. This is because the total energy loss of the $e^\pm$ pairs before losing most of its initial energy is roughly proportional to the distance they traveled, i.e., $\delta \varepsilon_{e^{\pm}}\propto l$ for $l\leq \lambda_{\rm IC}$, where $\lambda_{\rm IC}\approx 3m_e c^{2}/(4\gamma_{\rm e^\pm}\sigma_T u_{\rm cmb})$ is the cooling distance of the $e^\pm$ pairs and $u_{\rm cmb}$ is the energy density of the CMB radiation. The magnetic deflection angle reads $\theta_{\rm B}={l}/R_{\rm L}$, where $R_{\rm L}=\gamma_{\rm e^\pm}m_{\rm e}c^2/e B_{\rm IG}$ is the Larmor radius of the electrons. On the other hand, the arrival time of the inverse Compton GeV radiation is proportional to $l \theta_{\rm B}^2$,
i.e., $t \propto l^3$. Hence the observed flux is proportional to $d(\delta \varepsilon_{e^{\pm}})/dt\propto t^{-2/3}$.

\subsubsection{The secondary positron synchrotron radiation model}
In the external forward shock emitting region, ultra-relativistic positrons can be produced in the photomeson interaction between the X-ray photons and the ultra-energetic protons. The possibility of the photomeson interaction can be estimated as $\tau_{p\gamma} \sim 1.2\times 10^{-5}L_{\rm x,48.3}R_{17.5}^{-1}\Gamma_2^{-2}({\varepsilon_{\rm x,peak}\over 10~{\rm keV}})^{-0.3}({\varepsilon_{\rm x}\over 10~{\rm keV}})^{-0.7}(1+z)^{-1}$, where the X-ray spectrum $F_\nu \propto \nu^{-0.7}$ has been taken into account, $L_{\rm x}\propto t^{-1.35}$ for $t>10^{3}$ s is the luminosity of the X-ray emission and where $\varepsilon_{\rm x,peak}$ is the peak energy of the X-ray emission and $\varepsilon_{\rm x}$ is the energy of the X-ray photon. The energy of the protons should be $\varepsilon_{p} \approx 1.2\times 10^{17}~{\rm eV}~\Gamma_2^{2}(\varepsilon_{\rm x}/10~{\rm keV})^{-1}$ and the resulting positrons have an initial energy $\varepsilon_{e^{+}} \approx 6\times 10^{15}~{\rm eV}~\Gamma_2^{2}(\varepsilon_{\rm x}/10~{\rm keV})^{-1}$. Their synchrotron radiation in the forward shock region can generate $\sim 10~(\varepsilon_{\rm x}/10~{\rm keV})^{-2}$ GeV synchrotron radiation. Therefore this kind of process can give rise to GeV-TeV synchrotron radiation.
It is straightforward to show that the total energy of the positrons produced in the photomeson interaction can be estimated as $L_{e^{+}} \propto \int^{\varepsilon_{\rm p,II}}_{\varepsilon_{\rm p,I}}A_0 \gamma_{\rm p}^{1-p}\tau_{p\gamma}d\gamma_{\rm p} \propto \Gamma^{-p}R^{-1}L_{\rm x}/t$, where $A_0=(p-2)\epsilon_{\rm p}\gamma_{\rm p,m}^{p-2}E_{\rm k}/t$, $\varepsilon_{\rm p,I}\approx 1.2\times 10^{17}~{\rm eV}~\Gamma_2^{2}$, $\varepsilon_{\rm p,II}\approx 1.2\times 10^{19}~{\rm eV}~\Gamma_2^{2}$  and $\tau_{p \gamma}\propto \gamma_{\rm p}^{0.7}\Gamma^{-3.4}L_{\rm x}R^{-1}(1+z)^{-1}$ have been taken into account. Since $\Gamma\propto t^{-\frac{3}{8}, -\frac{1}{4}}$ and $R\propto t^{\frac{1}{4},\frac{1}{2}}$ for ISM and wind medium models respectively, $L_{e^{+}}$ drops with time more quickly than $t^{-1.8}$, at odds with the data.

\bigskip

\subsection{GeV-TeV emission powered by the forward shock electrons inverse-Compton-scattering off prompt photons}
For GRB 130427A the prompt X-ray emission was very strong and lasted a few hundred seconds (see Fig.\ref{fig:130427A-XRT}). Simultaneously, the ultra-relativistic GRB outflow drives energetic blast wave and accelerates a large amount of electrons. Some prompt photons will be up-scattered by the shock-accelerated electrons and get boosted to GeV-TeV energies when cross the forward (possibly also reverse) shock region(s). The resulting high energy $\gamma-$rays account for part of the observed GeV-TeV emission.

For illustration here we only calculate the high energy emission resulting in the external-inverse-Compton (EIC) scattering process in the second episode of the prompt emission ranging from $\sim 120$ s to $\sim 260$ s \citep{Golenetskii2013}. This is because at such a relatively late time the deceleration of the GRB outflow is most likely in the Blandford$-$McKee self-similar regime (Blandford \& McKee 1976) and the forward shock cooling/emission can be calculated in the standard way \citep{Piran99}.
We'd like to also point out that the very different temporal behaviors of the GeV and X-ray/soft $\gamma-$ray emission in such a time interval suggest that these emission are not from the same region.

The inverse Compton scattering is efficient if it is in the Thompson regime, requiring that in the rest frame of the electron the seed photon has an energy smaller than $m_{\rm e}c^{2}$, i.e., the random Lorentz factor of the electrons should not be higher than
\begin{eqnarray}
\gamma_{\rm s} &\sim & \Gamma m_{\rm e} c^2/4\epsilon_{\rm s},
\end{eqnarray}
where $\epsilon_{\rm s}$ is the energy of the seed photon (prompt photon). Therefore, most electrons with a random Lorentz factor $\sim \gamma_{\rm m}$ are cooled by the prompt photons at energies
\[\epsilon_{\rm s}\leq 4.3~{\rm keV}[{6(p-2)\over (p-1)}]^{-1}\epsilon_{\rm e,-1}^{-1}.\] {\it Such a $\epsilon_{\rm s}$ seems to be well below the peak energy $\sim 240 ~{\rm keV}$ of the prompt emission in the second episode} \citep{Golenetskii2013}. However the total energy released in the energy $<\epsilon_{\rm s}\sim 4.3~{\rm keV}$ is not ignorable since the current prompt emission spectrum is much softer than the typical ones $F_\nu \propto \nu^0$. With the observed spectrum $F_\nu \propto \nu^{-0.6}$ \citep{Golenetskii2013}, we find out that $\sim 1/5$ of the total energy was released below $\epsilon_{\rm s}$, i.e., ${\cal F}_{<\epsilon_{\rm s}}\sim {\cal F}/4 \sim 2.3\times 10^{-5}~{\rm erg~cm^{-2}}$, where ${\cal F}\sim 9\times 10^{-5}~{\rm erg~cm^{-2}}$ is the $20-1200$ keV energy fluence. The corresponding time-averaged luminosity is $L_{{<\epsilon_{\rm s}}}\sim 5\times 10^{49}$ erg/s, consistent with the {\it Swift} data (see Fig.\ref{fig:130427A-XRT}). Below we estimate the importance of the cooling of forward shock electrons by prompt emission.

In the rest frame of the shocked interstellar medium, the energy density of the seed photons can be estimated as
$U_\gamma \sim  L_{{<\epsilon_{\rm s}}}/4\pi R^2 \Gamma^2 c$.
The comoving energy density of the shock-generated magnetic field is
$U_{\rm B} \sim 4\Gamma^2 \epsilon_{\rm B} n m_{\rm p}c^2$ for ISM or $\sim 4\Gamma^2 \epsilon_{\rm B} n_{\rm w} m_{\rm p}c^2$ for wind medium.
The importance of the inverse Compton cooling caused by the prompt emission is given by the dimensionless parameter (Fan \& Piran 2006)
\begin{eqnarray}
{\cal Y} &=& U_\gamma/U_{\rm B} \nonumber\\
&\sim & \left\{%
\begin{array}{ll}
    0.22~L_{{<\epsilon_{\rm s}},50} E_{\rm k,54}^{-1}\epsilon_{\rm B,-2}^{-1} t_2({1+z\over 1.34})^{-1}, & \hbox{ISM;} \\
     1~L_{{<\epsilon_{\rm s}},50} E_{\rm k,54}^{-1}\epsilon_{\rm B,-2}^{-1} t_2 ({1+z\over 1.34})^{-1}, & \hbox{wind.} \\
\end{array}%
\label{eq:Y-1}
\right.
\end{eqnarray}
Such a ${\cal Y}$ seems to suggest a not very important inverse Compton cooling effect. However, as demonstrated below, intriguing radiation is  expected.

The number of high energy $\gamma-$rays generated by the forward shock electrons inverse-Compton-scattering off prompt emission can be straightforwardly estimated (e.g., Fan et al. 2005b, Gao et al. 2009). The possibility of one seed
photon being up-scattered (i.e., the optical depth) in the forward
shock region can be estimated as (Fan \& Piran 2006)
\begin{eqnarray}
\tau \sim  \left\{%
\begin{array}{ll}
    1.4\times 10^{-9}~E_{\rm k,54}^{1/4}n_{-2}^{3/4}t_2^{1/4}({1+z\over 1.34})^{-1/4}, & \hbox{ISM;} \\
     6\times 10^{-9}~E_{\rm k,54}^{-1/2}A_{*,-2}^{3/2}t_2^{-1/2}({1+z\over 1.34})^{1/2}, & \hbox{wind.} \\
\end{array}%
\label{eq:Y-1}
\right.
\end{eqnarray}
The total number of seed photons is
\begin{equation}
N_{\rm seed} \sim {\cal F}_{<\epsilon_{\rm s}}/<\epsilon_{\rm s}> \sim 2.5\times 10^{4}~{\rm cm^{-2}},
\end{equation}
where $<\epsilon_{\rm s}> \sim 0.6~{\rm keV}$ is the averaged energy of the seed photons within the energy range of $\sim 0.1~{\rm keV}-4.3~{\rm keV}$ for a spectrum $F_\nu \propto \nu^{-0.6}$. The total number of high energy photons detectable for Fermi with an effective area $S\sim 10^{4}~{\rm cm^2}$ is\footnote{Liu et al. (2013) argued that $n\sim 1~{\rm cm^{-3}}$. For such a large $n$, we have $N_{\rm \gamma,EIC} \sim 10$ and almost all the $>10$ GeV photons detected in the time interval of $120-260$ s can be accounted for. Moreover, the EIC emission likely plays a dominant role in producing GeV$-$TeV afterglow emission up to $t\sim 0.6$ day since in Liu et al. (2013)'s modeling the early ($t\sim 0.6$ day) strong soft X-ray emission should be mainly powered by the prolonged activity of central engine.}
\begin{eqnarray}
N_{\rm \gamma,EIC} \sim \tau N_{\rm seed} S \sim  \left\{%
\begin{array}{ll}
    0.3, & ~~\hbox{ISM;} \\
     1.5, & ~~\hbox{wind.} \\
\end{array}%
\label{eq:N-EIC}
\right.
\end{eqnarray}
The typical energy of the generated high energy $\gamma-$rays is expected to be
\begin{equation}
\epsilon_{\rm EIC} \sim \min\{ \gamma_{\rm m}^2,~\gamma_{\rm c}^2 \}<\epsilon_{\rm s}>\sim 24~{\rm GeV},
\end{equation}
where the cooling Lorentz factor of the forward shock electrons reads
\begin{eqnarray}
\gamma_{\rm c,4} \sim \left\{%
\begin{array}{ll}
    4.1~E_{\rm k,54}^{-{3\over 8}}\epsilon_{\rm B,-2}^{-1}n_{-2}^{-5/8}t_2^{1\over 8}({1+z\over 1.34})^{-{1\over 8}}({1+Y\over 2})^{-1}, & \hbox{ISM;} \\
    1.9~E_{\rm k,54}^{1\over 4}\epsilon_{\rm B,-2}^{-1}A_{*,-2}^{-5/4}t_2^{3\over 4}({1+z\over 1.34})^{-{3\over 4}}({1+Y\over 2})^{-1}, & \hbox{wind.} \\
\end{array}%
\label{eq:N-EIC}
\right.
\end{eqnarray}
and $Y$ is the inverse Compton parameter (including both the synchrotron-self Compton and the EIC radiation).

For GRB 130427A at a redshift $z=0.34$, the optical depth of the universe for 300 GeV-like $\gamma$-rays from interactions
with photons of the intergalactic background light is expected to be $\sim 1$ \citep{Gilmore2012}. Therefore whether the tens-GeV photons can be detected or not mainly depends upon their chance of escaping the emitting region. With eq.(13) of \citet{Zou2011} it is straightforward to show that even for 300~GeV photons the optical depth caused by the overlapping of the prompt photons with the forward shock region is $\sim 2.5\times 10^{-3}$, which is so small that can be ignored. Hence we conclude that the resulting tens-GeV photons can reach us.

Interestingly, in the time interval $140~{\rm s}<t<260~{\rm s}$ in coincidence with the second episode of the prompt emission, five afterglow photons at energies above 10 GeV have been recorded. Though the synchrotron-self-Compton origin of such photons can not be ruled out considering the somewhat small ${\cal Y}$, the EIC scattering origin due to the overlapping of prompt emission with the forward shock for such a tens-GeV-afterglow-emission enhancement is possible. As shown in Fig.12 of \citet{Fan08}, the duration of the EIC high-energy emission can be significantly
longer than the duration of the seed photon pulse
because the duration of the EIC emission
is affected by the spherical curvature of the blast wave
\citep{Beloborodov2005,Wang2006} and is further extended by the highly anisotropic
radiation of the up-scattered photons \citep{FP06}. Such an effect accounts for the fact that most of the $>10$ GeV afterglow photons detected in the time interval of $120-260$ s arrive after the main pulse of the BAT/XRT lightcurve peaking at $t\sim 120$ s (see Fig.1).

\section{A Possible Unified model for the prompt soft gamma-ray, optical and GeV emission of GRB 130427A, GRB 080319B and GRB 090902B}
Instead of proposing a model dedicated to fit the data of prompt emission of GRB 130427A, we try to outline a unified scenario to understand GRB 130427A, GRB 080319B and GRB 090902B together, motivated by the similarities displayed in the observational features summarized in Tab.\ref{Tab:cousins}. Please note that $\alpha_{\rm Band}$ and $\beta_{\rm Band}$ are the low and high energy spectral indexes of the GRBs fitted by the Band function (Band et al. 1992) and $E_{\rm p}$ is the observed peak energy of the spectrum.

\begin{table*}
\caption{General features of GRB 080319B, GRB 090902B and GRB 130427A.}
\begin{tabular}{llll}
\hline
Quantity & ~~GRB 080319B & ~~GRB 090902B & ~~GRB 130427A \\
\hline
$\alpha_{\rm Band}$ & ~~$0.833\pm 0.014$ & ~~$0.61\pm 0.01$ & ~~$0.789\pm 0.003$ $^a$\\
$\beta_{\rm Band}$ & ~~$3.499\pm 0.364$ &  ~~$3.8\pm 0.25$ & ~~$3.06\pm 0.06$ $^a$\\
$E_{\rm p}$ & ~~$651\pm 15$ keV & ~~$726\pm 8$ keV  &  ~~ $830\pm 5$ keV $^a$\\
$z$ & ~~~~0.937 & ~~~~1.822 &~~~~ 0.3399 \\
$E_{\rm \gamma,iso}$ & ~~$1.3\times 10^{54}$ erg & ~~$4\times 10^{54}$ erg & ~$8.5\times 10^{53}$ erg \\
Duration of prompt emission &~~~~$57$ s & ~~~~$26$ s &  ~~~~$\sim 138$ s $^b$\\
prompt optical emission  & ~~$\sim$ 20 Jy & ~~no observation & ~$\sim 4$ Jy \\
prompt GeV emission & ~~no observation & ~~$\sim 10^{-4}~{\rm erg~cm^{-2}}$ & ~~$\sim 10^{-4}~{\rm erg~cm^{-2}}$\\
main references & ~~~~1 & ~~~~2,3 &~~~~4,5,6,7 \\
\hline
\end{tabular}
\label{Tab:cousins}\\
\begin{minipage}{18cm}
$^a$ The time-averaged spectrum of the main phase of the burst
(from $T_0+0.002$ s to $T_0+18.432$ s) measured by Fermi-GBM \citep{Kienlin2013}.\\
$^b$ Most of the energy was released in the first $\sim 18$ s. \\
---References:
(1)~\citealt{Racu08}; (2)~\citealt{Palma09}; (3)~\citealt{Cucchiara09}; (4) \citealt{Kienlin2013}; (5)~\citealt{Zhu2013b}; (6)~\citealt{Golenetskii2013}; (7)~\citealt{Wren2013}.
\end{minipage}
\end{table*}

\emph{Prompt emission from the photosphere.} Prominent thermal radiation components have been identified in the prompt soft-$\gamma$ ray emission of GRB 090902B (e.g., Ryde et al. 2010; Zhang et al. 2011), which is the smoking-gun signature of the photospheric radiation. For GRB 080319B, some people tried to interpret both the ultra-strong soft $\gamma-$ray emission and the naked-eye optical flash within the internal shock scenarios \citep[e.g.,][]{KP08,Li2008,Yu2008}. However, the model of that the prompt optical and soft $\gamma$-ray emission are, respectively, the synchrotron and the
first-order inverse Compton radiation components of the internal
shocks is found to be disfavored \citep{Piran2009}. Moreover, the tight correlation $\Gamma \propto L^{0.3}$ found in the data analysis of GRBs \citep{Lv2012,Fan2012} predicts an
extremely low internal shock efficiency unless the slow material
shell has a width much wider than that of the fast shell, at odds with the data, where $L$ is the total luminosity of the outflow. Therefore we suggest that the internal shock origin of the prompt soft $\gamma-$ray emission is less likely. One attractive alternative is the so-called photospheric radiation model, in which the GRB prompt emission is mainly from the photosphere but suffers significant
modification and its spectrum is normally no longer thermal-like
(e.g., Rees \& M\'{e}sz\'{a}ros 2005; Beloborodov
2010; Lazzati et al. 2011). We adopt such a kind of model for the prompt emission of GRB 080319B and GRB 130427A, and investigate below whether strong prompt optical and GeV emission can be generated.

\emph{Bright optical flash from the synchrotron radiation of internal shocks.} If the prompt emission has a photospheric origin, the internal shocks are likely sub-relativistic since the contrast between the Lorentz factor of the shells is just by a factor of $\sim 2$. We denote the bulk Lorentz factor of the fast and slow shells as $\Gamma_{\rm f}$ and $\Gamma_{\rm s}$, respectively. The merged shell has a bulk Lorentz factor $\Gamma_{\rm i}$, which is between $\Gamma_{\rm s}$ and $\Gamma_{\rm f}$. Therefore the strength of the internal shock satisfies $\gamma_{\rm in}<(\Gamma_{\rm f}/\Gamma_{\rm s}+\Gamma_{\rm s}/\Gamma_{\rm f})< 1.25$ for $\Gamma_{\rm f}\approx 2\Gamma_{\rm s}$.  The comoving strength of the magnetic field in the emitting region can be estimated as $B_{\rm i}\sim 100~{\rm Gauss}~({3\epsilon_{\rm B,in}\over \epsilon_{\rm e,in}})^{1/2}L_{\rm in,52}^{1/2}R_{\rm i,16}^{-1}\Gamma_{\rm i,2.7}^{-1}$ \citep{FP08}, where $L_{\rm in}$ is the luminosity of the
internal shock radiation, and $\epsilon_{\rm e,in}$ and $\epsilon_{\rm B,in}$ are the fractions of the internal shock energy given to the electrons and magnetic field, respectively. The typical synchrotron radiation frequency of the internal shock electrons is $\nu_{\rm m} \sim 2.8\times 10^{6}~{\rm Hz}~\gamma_{\rm m,in}^{2}\Gamma_{\rm i}B_{\rm i}/(1+z) \sim 2.2\times 10^{14}(\gamma_{\rm m,in}/40)^{2}(3\epsilon_{\rm B}/\epsilon_{\rm e})^{1/2}L_{\rm in,52}^{1/2}R_{\rm i,16}^{-1}/(1+z)$ Hz, where $\gamma_{\rm m,in} \sim 40~[6(p-2)/(p-1)](\epsilon_{\rm e}/0.5)[(\gamma_{\rm in}-1)/0.2]$.
Following the standard treatment, the synchrotron self-absorption frequency can be estimated as $\nu_{\rm a} \sim  1.3\times 10^{15}~{\rm Hz}~({3\epsilon_{\rm B,in}\over \epsilon_{\rm e,in}})^{1/2}L_{53}^{2/(p+4)}L_{\rm in,52}^{p+2\over 2(p+4)}({\gamma_{\rm m,in}\over 40})^{2(p-1)\over p+4}\Gamma_{\rm i,2.7}^{-{2(p+6)\over p+4}}({\delta t \over 0.5~{\rm s}})$, i.e., above the optical band and then the optical emission is somewhat suppressed. The internal-shock-electrons with random Lorentz factor $\gamma_{\rm e}\leq \gamma_{\rm e,kn}\equiv \Gamma_{\rm i}m_{\rm e}c^{2}/[(1+z)E_{\rm p}] \sim 250 \Gamma_{\rm i,2.7}[(1+z)E_{\rm p}/1~{\rm MeV}]^{-1}$ are mainly cooled by the prompt soft $\gamma-$rays (i.e., the EIC process) and the cooling Lorentz factor can be estimated as
$\gamma_{\rm c,in} \sim 6 R_{\rm i,16}\Gamma_{\rm i,2.7}^3L_{\gamma,53}^{-1}$ \citep{FP08}, where $L_\gamma$ is the luminosity of the prompt soft $\gamma-$ray emission. So the comoving temperature of the
emitting region is $kT_{\rm i} \sim \min\{\gamma_{\rm m,in},~\gamma_{\rm c,in}\}m_{\rm e}c^{2}$ and the prompt optical flux density can be estimated as \citep{Zou09}
\begin{eqnarray}
f_{\nu_{\rm opt}} &\sim & {2\pi \nu_{\rm opt}^{2}(1+z)^{3}\Gamma_{\rm i}kT_{\rm i}\over c^{2}}({R_{\rm i} \over \Gamma_{\rm i} D_{\rm L}})^{2} \nonumber\\
&\sim & 3.4~{\rm Jy}~\nu_{\rm opt,14.7}^{2} \Gamma_{\rm i,2.7}^{-1}R_{\rm i,16}^{2}\nonumber\\
&& ({\min\{\gamma_{\rm m,in},~\gamma_{\rm c,in}\}\over 6})({1+z\over 2})^{3}D_{\rm L,28.3}^{-2}.
\end{eqnarray}
For reasonable parameters of GRB 080319B and GRB 130427A (i.e., $\Gamma_{\rm i}\sim 500-1000$ and $R_{\rm i} \sim 10^{16}$ cm), very bright optical flashes are expected, consistent with the data.

\emph{Energetic GeV emission from the EIC radiation of internal shocks.} As a result of the overlapping of the prompt emission and the optical radiation region, the electrons will scatter off the prompt emission and then produce high energy emission with a luminosity $\sim L_{\rm in}$ (Beloborodov 2005; Zou et al. 2009). The EIC radiation flux peaks at $\sim \min\{\gamma_{\rm m}^2,~\gamma_{\rm c}^{2}\} E_{\rm p} \sim 100~{\rm MeV}$ and the spectrum $F_\nu \propto \nu^{-p/2}$ can extend up to the energy $\sim (\Gamma_{\rm i} m_{\rm e} c^{2})^{2}/[(1+z)^{2}E_{\rm p}] \sim 0.25~{\rm TeV}~\Gamma_{\rm i,3}^{2}(E_{\rm p}/1~{\rm MeV})^{-1}(1+z)^{-2}$, as observed in GRB 090902B and GRB 130427A. Adopting eq.(13) of Zou et al. (2011), it is straightforward to show that the tens-GeV photons can escape without being significantly absorbed by the prompt $\gamma-$rays.

\section{Summary}
GRB 130427A was simultaneously detected by six $\gamma$-ray space telescopes and by three RAPTOR full-sky persistent monitors.  The isotropic-equivalent energy of the prompt emission is $\sim 10^{54}$ erg, rendering it the most powerful low-redshift ($z<0.5$) GRB detected so far (see Tab.\ref{Tab:low-z}). At a redshift of 0.3399, the very high energy emission ($\leq 300$ GeV) from GRB 130427A will not be considerably attenuated by the cosmic infrared/optical background. Together with the fact this nearby GRB is super-luminous, it is very favorable to detect the very high energy emission (Xue et al. 2009). The detection of four photons above 40 GeV (two above 70 GeV) by Fermi-LAT is in agreement with such a speculation.

The emission above 100 MeV lasted about one day (see Fig.2). As demonstrated by Zou et al. (2009), for bursts as energetic as GRB 080319B, the forward shock synchrotron radiation may be the dominant component of the afterglow emission below $\sim ~{\rm 10}$ GeV while the inverse Compton radiation mainly contributes at higher energies (see Kumar \& Barniol Duran 2009; Gao et al. 2009; Ghisellini et al. 2010 for interpreting the Fermi-LAT GeV afterglow data with the synchrotron radiation model). Such a conclusion seems to hold for GRB 130427A as well (see Fig.4 for numerical examples). In particular, for some photons at energies of tens-GeV, the forward shock synchrotron radiation model has been convincingly ruled out (see Fig.3) and an inverse Compton radiation origin is needed (see section 3.2 for discussion on alternative models). We also find out that the external-inverse-Compton-scattering of the prompt emission (the second episode, i.e., $t\sim 120-260$ s) by the forward-shock-accelerated electrons is expected to produce a few $\gamma-$rays at energies of tens-GeV, which may account for some $\gamma-$rays at energies $>10$ GeV detected in the same time interval.

We have also outlined a possible unified model for the prompt soft $\gamma-$ray, optical and GeV emission of GRB 130427A, GRB 080319B and GRB 090902B. In such a model the prompt soft $\gamma-$rays are mainly the photospheric radiation, while the subsequent internal shocks produce bright optical flash via synchrotron radiation and energetic GeV flash via the EIC scattering (see Section 4)\footnote{If the prompt emission of some bursts is triggered by the large scale magnetic energy dissipation, the subsequent (mildly-magnetized) internal shocks may also produce bright linearly-polarized optical flares as well as energetic GeV flash.}.

The IceCube collaboration reported their null detection of $>1$ TeV neutrinos in spatial and temporal
coincidence with GRB 130427A \citep{IceCube2013}. Such a result is a bit disappointed but not unexpected. For example, even in the internal shock model that is most favorable for producing PeV neutrinos, no detectable neutrino is expected if the proton spectrum is as soft as the electron spectrum (i.e., $dn/d\epsilon \propto \epsilon^{-4}$, as inferred from the prompt MeV emission). Only for the proton spectrum as hard as $dn/d\epsilon \propto \epsilon^{-2}$ and the kinetic energy of protons is about 10 times that of electrons, significant detection (i.e., about one event at PeV energies) by IceCube is possible. The high energy prompt emission does suggest such a hard spectrum. However it is likely powered at a radius $R_{\rm i} \sim 10^{16}$ cm (see Section 4), which is too large for efficient pion production. The possible high radiation efficiency of GRB 130427A \citep{Laskar2013} further reduces the chance of detecting the associated high energy neutrinos. In the photospheric radiation model for the prompt MeV emission, the non-detection of the associated TeV neutrino emission may suggest the absence of significant proton acceleration in the physical processes modifying the photon spectrum.

Finally we'd like to mention that the detection of one LAT photon of energy $\sim 72$ GeV at $t\sim 18.6$ s after the Fermi-GBM trigger of GRB 130427A can also be used to constrain the possible variation of the speed of light arising from quantum gravity effects. However, the limit is weaker than that set by the detection of one 31 GeV photon at $t\sim 0.73$ s after the trigger of GRB 090510 \citep{Abdo09}.

\section*{Acknowledgments}
We thank the anonymous referee for the insightful comments/suggestions. We are also grateful to S. Kobayashi, Y. C. Zou, L. Shao and D. Xu for helpful discussion and D. A. Kann for suggestions. This work was supported in part by 973 Program of China under grants 2009CB824800 and 2013CB837000, National Natural Science of China under grants 11173064, 11163003 and 11273063, and by China Postdoctoral science foundation under grant 2012M521137. YZF is also supported by the 100
Talents program of Chinese Academy of Sciences and the Foundation for
Distinguished Young Scholars of Jiangsu Province, China (No. BK2012047). PHT is supported by the National Science Council of the Republic of China (Taiwan) through grant NSC101-2112-M-007-022-MY3.

\clearpage


\begin{thebibliography}{}
\bibitem[Abdo et al. (2009a)]{Abdo09} Abdo, A. et al. 2009a, {Nature}, 462, 331

\bibitem[Abdo et al. (2009b)]{Palma09} Abdo A. et al. 2009b, {ApJ}, 706, L138

\bibitem[Abdo et al. (2013)]{Abdo2013} Abdo A. et al. (Fermi collaboration) 2013,
arXiv:1303.2908

\bibitem[Ackerman et al. (2013)]{Ackermann2013} Ackermann, M. et al. 2013, ApJ, 763, 71

\bibitem[Acciari et al. (2011)]{Acciari2011} Acciari, V.~A. et al. (VERITAS collaboration) 2011, ApJ, 732, 62

\bibitem[Aharonian et al. (2009)]{Aharonian09} Aharonian, F., et al. (H.E.S.S. collaboration) 2009, A\&A, 495, 505

\bibitem[Albert et al. (2007)]{Albert07} Albert, J., et al. (MAGIC collaboration) 2007, ApJ, 667, 358

\bibitem[Amati et al. (2007)]{Amati2007} Amati, L., Della Valle,M., Frontera, F., Malesani, D., Guidorzi, C., Montanari, E., \& Pian, E. 2007, A\&A, 463, 913


\bibitem[Band et al. (1993)]{Band93} Band D. et al., 1993, ApJ, { 413}, 281

\bibitem[Barthelmy et al. (2013)]{Barthelmy2013} Barthelmy, S. D. et al. 2013 GCN Circ. 14470 (http://gcn.gsfc.nasa.gov/gcn3/14470.gcn3)

\bibitem[Beloborodov (2005)]{Beloborodov2005} Beloborodov, A. M. 2005, ApJ, 618, L13

\bibitem[Beloborodov (2010)]{Beloborodov2010} Beloborodov, A. M. 2010, MNRAS, 407, 1033

\bibitem[Blandford \& McKee (1976)]{Blandford1976} Blandford, R. D., \& McKee, C. F., 1976, Phys. Fluids, 19, 1130

\bibitem[Blaufuss (2013)]{IceCube2013} Blaufuss, E. (IceCube collaboration) 2013, GCN Circ. 14520 (http://gcn.gsfc.nasa.gov/gcn3/14520.gcn3)

\bibitem[Butler et al.(2007)]{2007ApJ...671..656B} Butler, N.~R., Kocevski,
D., Bloom, J.~S., \& Curtis, J.~L.\ 2007, \apj, 671, 656

\bibitem[Cheng \& Wei (1996)]{CW96}Cheng, K.S., \& Wei, D.M., 1996, MNRAS, { 283}, L133

\bibitem[Cheng \& Cheng (1996)]{Cheng1996} Cheng, L. X., \& Cheng, K. S., 1996, ApJ, 459, L79

\bibitem[Cucchiara et al. (2009)]{Cucchiara09} Cucchiara, A., Fox, D. B., Tanvir, N., \& Berger, E., 2009, GCN 9873

\bibitem[Dai \& Lu (1998)]{Dai1998} Dai, Z. G., \& Lu, T. 1998, MNRAS, 298, 87

\bibitem[Dai \& Lu (2002)]{Dai2002} Dai, Z. G., \& Lu, T. 2002, ApJ, 580, 1013

\bibitem[Dermer et al. (2000)]{Dermer2001} Dermer, C. D., Chiang, J., \& Mitman, K. E., 2000, ApJ, 537, 785

\bibitem[Evans et al. (2009)]{Evans2009} Evans, P. A., et al. 2009, MNRAS, 397, 1177

\bibitem[Fan \& Piran (2006)]{FP06}
Fan, Y. Z. \& Piran, T.
2006, MNRAS, {370}, L24

\bibitem[Fan \& Piran (2008)]{FP08}
Fan, Y. Z. \& Piran, T.
2008, {Front. Phys. China.}, { 3}, 306

\bibitem[Fan et al. (2008)]{Fan08}
Fan, Y. Z., Piran, T., Narayan, R., \& Wei, D. M.
2008, MNRAS, 384, 1483

\bibitem[Fan et al. (2012)]{Fan2012} Fan, Y. Z., Wei, D. M., Zhang, F. W., \& Zhang, B. B.
2012, ApJL, 755, L6

\bibitem[Fan et al. (2005a)]{Fan05a} Fan, Y. Z., Zhang, B., \& Wei, D. M. 2005a, ApJ, 629, 334

\bibitem[Fan et al. (2005b)]{Fan05b} Fan, Y. Z., Zhang, B., \& Wei, D. M. 2005b, MNRAS, 361, 965

\bibitem[Flores et al. (2013)]{Flores2013} Flores, H. et al., 2013, GCN Circ. 14491 (http://gcn.gsfc.nasa.gov/gcn3/14491.gcn3)

\bibitem[Gao et al. (2009)]{Gao2009} Gao, W. H., Mao, J. R., Xu, D., \& Fan, Y. Z., 2009, ApJ, 706, L33

\bibitem[Ghisellini et al. (2010)]{Ghisellini2010} Ghisellini, G., Ghirlanda, G., Nava, L., \& Celotti, A. 2010, MNRAS, 403, 926

\bibitem[Gilmore et al. (2012)]{Gilmore2012} Gilmore, R., Somerville, R., Primack, J., \& Dom\'{i}nguez, A. 2012, MNRAS, 422, 3189

\bibitem[Golenetskii et al. (2013)]{Golenetskii2013} Golenetskii, S., et al. 2013, GCN Circ. 14487 (http://gcn.gsfc.nasa.gov/gcn3/14487.gcn3)

\bibitem[Goodman (1986)]{Good86}Goodman J., 1998, ApJ, { 308}, L47

\bibitem[He et al. (2011)]{He2011} He, H. N., Wu, X. F., Toma, K., Wang, X. Y., \& M\'{e}sz\'{a}ros, P. 2011, ApJ, 733,
22

\bibitem[Holland et al. (2010)]{Holland2010} Holland, S. T., et al. 2010, ApJ, 717, 223

\bibitem[Horan (2007)]{Horan07} Horan, D., et al. 2007, ApJ, 655, 396

\bibitem[Hurley et al. (1994)]{Hurley94}  Hurley, K. et al. 1994, Nature, 372, 652

\bibitem[Jarvis et al. (2010)]{Jarvis2010} Jarvis, A. et al. 2010, ApJ, 722, 862


\bibitem[Kumar \& Panaitescu (2008)]{KP08} Kumar, P., \&  Panaitescu, A. 2008, MNRAS, { 391}, L19

\bibitem[Kumar et al. (2009)]{Kumar2009} Kumar, P., \& Barniol Duran, R. 2009, MNRAS, 400, L75

\bibitem[Laskar et al. (2013)]{Laskar2013} Laskar, T. et al. 2013, ApJ submitted (arXiv:1305.2453)

\bibitem[Lazzati et al. (2011)]{Lazzati2011} Lazzati, D., Morsony, B. J., \& Begelman, M. C. 2011, ApJ, 732, 34

\bibitem[Lennarz \& Taboada (2013)]{Lennarz2013} Lennarz, D., \& Taboada, I. 2013, GCN Circ. 14549 (http://gcn.gsfc.nasa.gov/gcn3/14549.gcn3)

\bibitem[Levan et al. (2013)]{Levan2013} Levan, A. J., Cenko, S. B., Perley, D. A., \& Tanvir, N. R. 2013, GCN Circ. 14455
(http://gcn.gsfc.nasa.gov/gcn3/14455.gcn3)

\bibitem[Li \& Waxman (2008)]{Li2008} Li, Z.,\& Waxman, E., 2008, ApJ, 674, L65

\bibitem[Liu et al. (2013)]{Liu2013} Liu, R. Y., Wang, X. Y., \& Wu, X. F., 2013, ApJL submitted (arXiv:1306.5207)

\bibitem[L\"{u} et al. (2012)]{Lv2012} L\"{u}, J. et al. 2012, ApJ, 751, 49

\bibitem[M\'esz\'aros (2002)]{Mesz02}
M\'esz\'aros, P.
2002, {Ann. Rev. Astron. Astrophy.}, { 40}, 137

\bibitem[Maselli et al. (2013)]{Maselli2013} Maselli, A., Beardmore, A. P.,
Lien, A. Y., Mangano, V.,
Mountford, C. J., Page, K. L., Palmer, D. M., \&  Siegel, M. H. 2013, GCN Circ. 14448 (http://gcn.gsfc.nasa.gov/gcn3/14448.gcn3)

\bibitem[Perley et al. (2013)]{Perley2013}
Perley, D. A. et al. 2013, ApJ submitted
(arXiv:1307.4401)

\bibitem[Piran (1999)]{Piran99}
Piran, T. 1999,
{Phys. Rep.}, { 314}, 575

\bibitem[Piran et al. (2009)]{Piran2009} Piran T., Sari R., Zou Y. C., 2009, MNRAS, 393, 1107

\bibitem[Plaga (1995)]{Plaga1995} Plaga, R. 1995, Nature, 374, 430

\bibitem[Pozanenko et al.(2013)]{Pozanenko2013} Pozanenko, A., Minaev, P., \& Volnova, A. 2013, GCN Circ. 144484 (http://gcn.gsfc.nasa.gov/gcn3/14484.gcn3)

\bibitem[Racusin et al. (2008)]{Racu08} Racusin, J. L., et al. 2008,
 {Nature}, { 455}, 183

\bibitem[Razzaque et al. (2010)]{Razzaque2010}
 Razzaque, S., Dermer, C. D., \& Finke, J. D.  2010,
The Open Astronomy Journal, 3, 150

\bibitem[Rees \& M\'esz\'aros (2005)]{RM2005}
Rees, M. J. \& M\'esz\'aros, P. 2005, ApJ, 628, 847

\bibitem[Ryde et al. (2010)]{Ryde10} Ryde, F., et al. 2010, ApJ, 709, L172

\bibitem[Sari \& Esin (2001)]{Sari2001} Sari, R., \& Esin, A. A., 2001, ApJ, 548, 787

\bibitem[Smith et al. (2013)]{Smith2013} Smith, D. M., Csillaghy, A.,
Hurley, K., Hudson, H., Boggs, S., \&
Inglis, A. 2013, GCN Circ. 14590 (http://gcn.gsfc.nasa.gov/gcn3/14590.gcn3)

\bibitem[Tam et al. (2012)]{Tam2012} Tam,P. H. T.,  Kong, A. K. H., \& Fan, Y. Z. 2012, ApJ, 754, 117

\bibitem[Tam et al. (2013)]{Tam2013} Tam,P. H. T. et al., 2013, ApJL, 771, L13 (arXiv:1305.3217)

\bibitem[Kienlin et al. (2013)]{Kienlin2013} von Kienlin, A., 2013, GCN Circ. 14473 (http://gcn.gsfc.nasa.gov/gcn3/14473.gcn3)

\bibitem[Verrecchia et al. (2013)]{Verrecchia2013} Verrecchia, F., et al. 2013, GCN Circ. 14515 (http://gcn.gsfc.nasa.gov/gcn3/14515.gcn3)

\bibitem[von Kienlin (2010)]{2010GCN..10381...1V}
von Kienlin A., 2010, GCN Circ., 10381, 1

\bibitem[Wang et al. (2006)]{Wang2006} Wang, X. Y., Li, Z., \& M\'{e}sz\'{a}ros, P., 2006, ApJ, 641, L89

\bibitem[Wren et al. (2013)]{Wren2013} Wren, J., Vestrand, W. T., Wozniak, P., \& Davis, H. 2013, GCN Circ. 14476 (http://gcn.gsfc.nasa.gov/gcn3/14476.gcn3)

\bibitem[Xu et al. (2013)]{Xu2013} Xu, D., et al. 2013, ApJL in press (arXiv:1305.6832)

\bibitem[Xue et al. (2009)]{Xue09} Xue, R. R., et al. 2009, ApJ, 703, 60

\bibitem[Yu et al. (2009)]{Yu2008} Yu, Y. W., Wang, X. Y., \& Dai, Z. G., 2009, ApJ, 692, 1662

\bibitem[Zhang \& M\'esz\'aros (2001)]{Zhang2001} Zhang, B., \& M\'esz\'aros, P., 2001, ApJ, 559, 110

\bibitem[Zhang \& M\'esz\'aros (2004)]{Zhang04} Zhang, B., \& M¡äesz¡äaros, P. 2004, Int. J. Mod. Phys. A, 19, 2385

\bibitem[Zhang et al. (2011)]{Zhang2011} Zhang, B. B., et al., 2011, ApJ, 730, 141

\bibitem[Zhang et al. (2012a)]{Zhang2012} Zhang, B. B., et al., 2012a, ApJ, 756, 190

\bibitem[Zhang et al. (2012b)]{Zhang2012a} Zhang, F. W., Shao, L., Yan, J. Z., \& Wei, D. M. 2012b, ApJ, 750, 88


\bibitem[Zhu et al. (2013a)]{Zhu2013} Zhu, S., Racusin, J., Kocevski, D., McEnery, J., Longo, F., Chiang, J., \& Vianello, G. 2013a, GCN Circ. 14471
(http://gcn.gsfc.nasa.gov/gcn3/14471.gcn3)

\bibitem[Zhu et al. (2013b)]{Zhu2013b} Zhu, S., Racusin, J., Kocevski, D., McEnery, J., Longo, F., Chiang, J., \& Vianello, G. 2013b, GCN Circ. 14508
(http://gcn.gsfc.nasa.gov/gcn3/14508.gcn3)

\bibitem[Zou et al. (2009)]{Zou09} Zou, Y.C., Fan, Y.Z., \& Piran, T., 2009, MNRAS, {396}, 1163

\bibitem[Zou et al. (2011)]{Zou2011} Zou, Y.C., Fan, Y.Z., \& Piran, T., 2011, ApJ, {726}, L2

\end{thebibliography}
\end{document}